\documentclass[prb,11pt]{revtex4-1}

\usepackage{amsmath}    
\usepackage{graphicx}   
\usepackage{verbatim}   
\usepackage{color}      
\usepackage{subfigure}  
\usepackage{hyperref}   
\usepackage[parfill]{parskip}                           
\usepackage{caption}                                    
\usepackage{amssymb}
\usepackage{amsmath}
\usepackage{graphicx}                                   
\usepackage{epstopdf}
\usepackage{siunitx}
\usepackage{mathtools}                                  
\usepackage{float}
\usepackage{version}
\usepackage{cases}                                              
\usepackage{bm}
\usepackage{pstricks}                                   
\usepackage{ulem}
\usepackage{mathrsfs}
\usepackage{tikz}
\usetikzlibrary{decorations.markings,arrows}


\begin{document}
\title{Universality and Thermodynamics of Turbulence}

\author{Damien Geneste$^{1}$, Hugues Faller$^{1}$*, Florian Nguyen$^{2}$, Vishwanath Shukla$^{3}$, Jean-Philippe Laval$^{2}$,  Francois Daviaud$^{1}$, Ewe-Wei Saw$^{1}$ and B{\'e}reng{\`e}re Dubrulle $^{1}$}

\affiliation{$^{1}$ \quad SPEC, CEA, CNRS, Université Paris-Saclay, CEA Saclay, Gif-sur-Yvette, France}
\affiliation{$^{2}$ \quad Laboratoire de Mécanique des Fluides de Lille, CNRS, ONERA, Centrale Lille, Univ. Lille, Arts et Metiers ParisTech, Kampé de Fériet, F-59000, Lille, France}
\affiliation{$^3$ Universit\'e C\^ote d'Azur, Institut de Physique de Nice, CNRS, Nice, France}

\affiliation{Correspondence: hugues.faller@normalesup.org; Tel.: +33-16908 3015}

\date{Friday, March 1 2019}

\begin{abstract}
We investigate universality of the Eulerian velocity structure functions using 
velocity fields obtained from the stereoscopic particle image velocimetry (SPIV) technique in experiments and the direct numerical simulations (DNS) of the Navier-Stokes equations.
 We show  that the numerical and experimental velocity structure functions up to order 9 follow a log-universality \cite{Castaing93}; we
 find that they collapse on a universal curve, if we use units that include logarithmic dependence on the Reynolds number. We then investigate the meaning and consequences of such log-universality, and show that it is connected with the properties of a "multifractal free energy", based on an analogy between multifractal and themodynamics. We show that in such a framework, the existence of a fluctuating dissipation scale is associated with a phase transition describing the relaminarisation of rough velocity fields with different H\"older exponents. Such a phase transition has been already observed using the Lagrangian velocity structure functions, but was so far believed to be out of reach for the Eulerian data.
\end{abstract}

\maketitle
\section{Introduction}
\label{Intro}
A well-known feature of any turbulent flow is the Kolmogorov-Richardson cascade by which energy is transferred from large to small length scales until the Kolmogorov length scales below which it is removed by viscous dissipation. This energy cascade is a non-linear and an out-of-equilibrium universal process. Moreover, the corresponding non-dimensional energy spectrum $E(k)/\epsilon^{2/3}\eta^{5/3}$ is an universal function of $k\eta$, where $\eta=(\nu^3/\epsilon)^{1/4}$ is the Kolmogorov length scale, $\epsilon$ the mean energy dissipation rate per unit mass, and $\nu$ the kinematic viscosity. However, there seem to be little dependences on the Reynolds number, boundary, isotropy or homogeneity conditions \cite{Dubrulle19}. In facts, the energy spectrum is  based upon a quantity, the velocity correlation, that is quadratic in velocity. Nevertheless, it is now well admitted that the universality does not carry over for statistical quantities that involve higher order moments. For example, the velocity structure functions of order $p$,  given by $S_p(\ell)=\langle \|u(\textbf{x}+\textbf{r})-u(\textbf{x})\|^p  \rangle_{\textbf{x},\|\textbf{r}\|=\ell}$ are not universal, at least when expressed in units of the Komogorov scale $\eta$ and velocity $u_{\mathrm{K}}=(\nu\epsilon)^{1/4}$(see below, section \ref{sec:K41} for an illustration).\

The mechanism behind this universality breaking was identified by \cite{[P85]}, wherein a generalization of the Kolmogorov theory introduced, based on the hypothesis that a turbulent flow is multifractal. In this model, the velocity field is characterized locally by an exponent $h$, such that $|\delta_\ell u(\textbf{x})|\equiv \langle\|\textbf{u}(\textbf{x}+\textbf{r})-\textbf{u}(\textbf{x})\|\rangle_{\|\textbf{r}\|=\ell}\sim \ell^{h(\textbf{x})}$; here $h$ is a stochastic function that follows a large deviation property \cite{EyinkLN} $\mathbb{P}\left(\log(|\delta_\ell u|/u_0)=h\log\left(\ell/L_0\right)\right)\sim \left(\ell/L_0\right)^{C(h)}$, where $u_0$ (resp. $L_0$) is the caracteristic integral velocity (resp. length), and $C(h)$ is the multifractal spectrum. Velocity fields with $h<1$ are rough in the limit $\ell\to 0$. In real  flows, any rough field with $h>-1$ can be regularized at sufficiently small scale (the "viscous scale") by viscosity. The first computation of such dissipative scale was performed by Paladin and Vulpiani \cite{[P87]}, who showed that it scales with viscosity like $\eta_h\sim \nu^{1/(1+h)}$, thereby generalizing the Kolmogorov scale, which corresponds to $h=1/3$. Such a dissipative scale fluctuates in space and time (along with $h$), resulting in non-universality for high order moments, at least when expressed in units of $\eta$ and $u_{\mathrm{K}}$.\

A few years later, Frisch and Vergassola \cite{Frisch_1991} claimed that the universality of the energy spectrum 
can be recovered, if the fluctuations of the dissipative length scale are taken into account by introducing a new non-dimensionalisation procedure. The new prediction is that $\log\left(E(k)\epsilon^{-\frac23}\eta^{-\frac53}\right)/\log(\text{Re})$ should be a universal function of $\log(k\eta)/\log(\text{Re})$, where $\text{Re}$ is the Reynolds number. This claim was examined by Gagne et al., later  using data from the Modane wind tunnel experiments \cite{Castaing93}. They further suggested that the prediction can be extended to the velocity structure functions, so that, at any given $p$ $\log(S_p(\ell)/u_{\mathrm{K}}^p)/\log(\text{Re})$ should be a universal function of $\log(\ell/\eta)/\log(\text{Re})$; they found good agreement for $p$ up to $6$.  
 The velocity measurements, in the above experiments, were performed using hot wire anemometry, which provide access to only one component of velocity.  To our knowledge, no further attempts have been made to check the claim with 
 more realistic measurements. 
 
 The purpose of the present paper is to reexamine this claim; however, now using the velocity fields obtained from the Stereoscopic  Particule Image Velocimetry (SPIV) in experiments and the direct numerical simulations (DNS) of the Navier-Stokes equations (NSE). We show  that the numerical and experimental velocity structure functions up to order 9 follow a log-universality \cite{Castaing93}; they indeed collapse on a universal curve, if we use units that include $\log(\text{Re})$ dependence. We then investigate the meaning and consequences of such a log-universality, and show that it is connected with the properties of a "multifractal free energy", based on an analogy between multifractal and thermodynamics (see \cite{[M91]} for summary). We show that in such a framework, the existence of a fluctuating dissipation length scale is associated with a phase transition describing the relaminarisation of rough velocity fields with different H\"older exponents.

\section{Experimental and numerical setup}
\subsection{Experimental facilities and parameters}
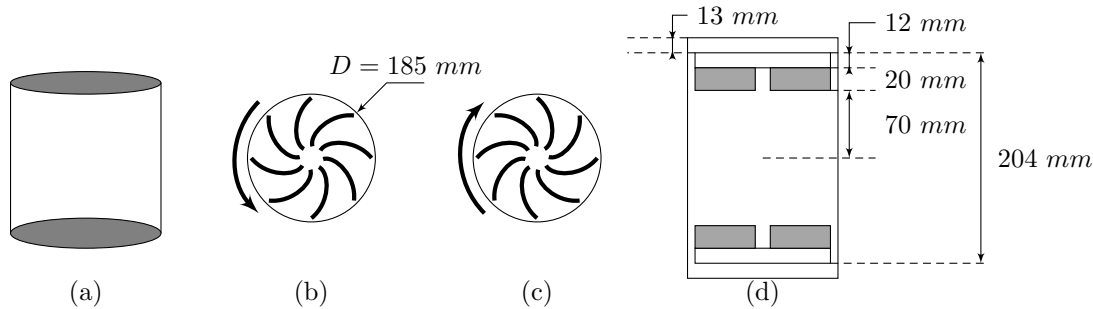
\begin{figure}
\centering
\begin{tikzpicture}
     \draw[fill=gray] (0,0) ellipse (1cm and 0.15cm);
     \draw (-1,0)--(-1,-2);
     \draw (1,0)--(1,-2);
     \draw[fill=gray] (0,-2) ellipse (1cm and 0.2 cm);   
     \draw (3,-1) circle (0.85cm);
     \draw[<-,>=latex'] (3.601,-0.399)--(4,0)--(4.5,0);
     \node[above] at (4.25,0) {$D= 185 \ mm$}; 
     \begin{scope}[shift={(3,-1)}]
     \foreach \r in {0,45,90,135,180,225,270,315}
     {\begin{scope}[rotate = \r]
        \draw[ultra thick] (0.12,0.1)..controls (0.2,0.25) and (0.6,0.25) .. (0.8,0);
     \end{scope}}
      \draw[->, >=latex', ultra thick] (-0.7,0.75) arc (135:225:1.05cm);
     \end{scope}  
     \draw (6,-1) circle (0.85cm);
     \begin{scope}[shift={(6,-1)}]
     \foreach \r in {0,45,90,135,180,225,270,315}
     {\begin{scope}[rotate = \r]
        \draw[ultra 
        thick] (0.12,-0.1)..controls (0.2,-0.25) and (0.6,-0.25) .. (0.8,0);
     \end{scope}}
      \draw[<-, >=latex', ultra thick] (-0.7,0.75) arc (135:225:1.05cm);
     \end{scope}
     
    \node at (0,-2.8) {(a)};
    \node at (3,-2.8) {(b)};
    \node at (6,-2.8) {(c)};
    \node at (9,-2.8) {(d)};
    \draw (8,-2.6) rectangle (10,0.6);
    \draw (8.1,0.4) rectangle (9.9,0.2);
    \draw[fill = gray!70] (8.1,0.2) rectangle (8.9,-0.1);
    \draw[fill = gray!70] (9.9,0.2) rectangle (9.1,-0.1);
    \node[right] at (10.5,0.05) {$20 \ mm$};
    \draw[densely dashed] (9.9,0.4)--(12,0.4);
    \draw[densely dashed] (9.9,0.2)--(10.5,0.2);
    \draw[densely dashed] (9.9,-0.1)--(10.5,-0.1);
    \draw[densely dashed] (9,-1)--(10.5,-1);
    \draw[densely dashed] (9.9,-2.4)--(12,-2.4);
    \draw (8.1,-2.4) rectangle (9.9,-2.2);
    \draw[fill = gray!70] (8.1,-2.2) rectangle (8.9,-1.9);
    \draw[fill = gray!70] (9.9,-2.2) rectangle (9.1,-1.9);
    \draw[>-<,>=latex'] (10.15,0.5)--(10.15,0.1);
    \draw (10.15,0.4)--(10.15,0.8)--(10.5,0.8) node[right] {$12 \ mm$};
    \draw[<->,>=latex'] (10.15,-0.1)--(10.15,-1);
    \node[right] at (10.5,-0.55) {$70 \ mm$};
    \draw[<->,>=latex'] (11.9,0.4)--(11.9,-2.4);
    \node[right] at (12,-1) {$204 \ mm$};
	\draw[densely dashed] (8,0.6)--(7.2,0.6);
	\draw[densely dashed] (8.1,0.4)--(7.2,0.4);
	\draw[>-<,>=latex'] (7.8,0.3)--(7.8,0.7);
	\draw (7.8,0.6)--(7.8,0.9)--(8,0.9) node[right] {$13 \ mm$};
\end{tikzpicture}   
\caption{Von Kármán swirling flow generator. (a) normal view, bottom (b) and top (c) impellers rotating -both seen from the center of the cylinder, and (d) sketch with the relevant measures. A device not shown here maintains the temperature constant during the experiment. Both impellers are counter-rotating.}
\label{fig:sketch_VK}
\end{figure}

We use  experimental  velocity field  described in \cite{saw_debue_kuzzay_daviaud_dubrulle_2018}. The radial, axial and azimuthal velocity are measured in a von Kármán flow, using Stereoscopic  Particule Image Velocimetry technique at different resolutions $\Delta x$. The von Kármán flow is generated in a cylindrical tank of radius $R=\SI{10}{cm}$ through counter-rotation of two independent impellers with curved blades. The flow was maintained in a turbulent state at high Reynolds number by two independent impellers, rotating at frequency $F$.  Figure \ref{fig:sketch_VK} shows the sketch of the experimental setup. The five experiments 
were performed in conditions so that the non-dimensional mean energy dissipation per unit mass is constant. The viscosity was monitored using mixture of water and glycerol, so as to vary the Kolmogorov length $\eta$. Table \ref{tab:exp_parameter} summarizes the different parameters; $R_{\lambda}= \lambda u^{\mathrm{rms}}/\nu$ is the Reynolds number based on the Taylor length scale $ \lambda = \sqrt{\frac{ \langle\textbf{u}^2\rangle}{\langle\nabla \textbf{u}^2\rangle}}$, the mean squared velocity $u^{\mathrm{rms}}$ and the viscosity $\nu$. 

All velocity measurements are performed in a vertical plane that contains the rotation axis. The case A corresponds to measurements over the whole plane contained in between the two impellers, and extending from one side to the other side of the cylinder.   Its resolution is 5 to 10 times coarser than similar measurements performed by zooming on a region centered around the symmetry point of the experiment (on the rotation axis, half way in between the two impellers), over a square window of size $\SI{2}{cm}\times\SI{2}{cm}$.
   
\begin{table}
    \centering
\begin{tabular}{@{\hspace{4mm}}l| c c c c@{\hspace{4mm}} c@{\hspace{4mm}} l r c}
    \hline
    \hline
    Case & F (Hz) & Glycerol part & $\text{Re}$ & $R_\lambda$ & $\eta$ (mm) & $\Delta x$ &   Frames & Symbol  \\
    \hline 
    A  & 5 & 0\%   & $3\times10^{5}$ & $1,9 
    \times10^{3}$ & 0.02 & 2.4 & $3\times 10^4$ & $\circ$ \\
    B  & 5 & 0\%   &$3\times10^{5}$ & $2,7
    \times10^{3}$& 0.02 & 0.48 & $3\times 10^4$ &$\Box$\\
    C  & 5 & 0\%   &$3\times10^{5}$ & $2,5
    \times10^{3}$& 0.02 & 0.24 & $2\times 10^4$ &$\Diamond$\\
    D  & 1 & 0\%   &$4\times10^{4}$ & $9,2 
    \times10^{2}$& 0.08 & 0.48 & $1\times 10^4$ &$\vartriangle$\\
    E  & 1.2 & 59\% &$6\times10^{3}$ & $2,1 
    \times10^{2}$& 0.37 & 0.24 & $3\times 10^4$ & $\star$\\
    \hline
    \hline
\end{tabular}
    \caption{Parameters for the 5 experiments realized $(\mathbf{A},\mathbf{B},\mathbf{C},\mathbf{D} \ and \ \mathbf{E})$. F is the rotation frequency of the discs, $\text{Re}$ refers to the Reynolds number based on the diameter of the tank, $R_{\lambda}$ is the Reynolds based on the Taylor micro-scale. $\eta$ gives the estimated Kolmogorov length according to the experiment and $\Delta x $ refers to the spatial resolution of SPIV measurements. The last columns gives the number of frames used and the number of points over which are calculated the statistics. Except for (\textbf{E}), the Reynolds are much larger than those available with DNS. Table taken from \cite{saw_debue_kuzzay_daviaud_dubrulle_2018}}
    \label{tab:exp_parameter}
\end{table}
\subsection{Direct Numerical Simulation}
The direct numerical simulations (DNS), based on pseudo-spectral method, were performed in order to compare with our experimental data. The DNS runs with  $R_{\lambda} = 25$, $R_{\lambda} = 80$, $R_{\lambda} = 90$ and $R_{\lambda} = 138$ were performed 
using the NSE solver VIKSHOBHA~\cite{Debue18},  whereas the run with $R_{\lambda}=56$ was carried out using another independent pseudo-spectral method based NSE solver. The velocity field $\bm{u}$ was computed on a $2\pi$ triply-periodic box. \\
Turbulent flow in a statistically steady state was obtained by using the Taylor-Green type external forcing in the NSE at wavenumber $k_{f}=1$ and amplitude $f_{0}=0.12$, the value of viscosity was varied in order to obtain different values of $R_{\lambda}$ (see Ref.~\cite{Debue18} for more details).
\begin{table}
    \centering
\begin{tabular}{@{\hspace{4mm}}l l l c l l l l l@{\hspace{4mm}}}
    \hline
    \hline
    $R_\lambda$ & $\eta$ & $k_{\max}\eta$ & $N_{x}\! \times N_{y}\! \times \!N_{z}$ & $\ell_{\min}/\eta$ &$\widetilde{k_{f}}$& Samples &  Symbol  \\
    \hline 
    25  & 0.079 & 3.35 & $128^{3}$  & 0.635 & 1&  5000 &  $\star$ \\
    56  & 0.034  & 6.42 & $256^{3}$  &0.31 & 1 &  105 000&$\vartriangle $ \\
    80  & 0.020 & 1.68 & $256^{3}$  & 1.22 & 1&  270 000 & $\Box$ \\
    90  & 0.017 & 5.70 & $1024^{3}$ & 0.36 & 1 & 10 000 & $\Diamond$ \\
    138 &0.009 &1.55 & $512^{3}$ & 1.37 & 1 &12 000 & $\circ$\\
    \hline
    \hline
\end{tabular}
    \caption{Parameters for the DNS. $R_\lambda$ is  the Reynolds based on the Taylor micro-scale. $\eta$ is  the Kolmogorov length. The third column gives resolution of the simulation through $k_{\max}\eta$,where $k_{\max}=N/3$ is the maximum wavenumber. The fourth column gives the grid size; notice that the length of the box is $2\pi$. Here, $\ell_{\min}$ is the smallest scale available for the calculations of the wavelets.  $\tilde k_f$ is the forcing scale. The Sample columns gives the number of points (frames $\times$ gridsize) over which are calculated the statistics.}
    \label{tab:DNS_parameter}
\end{table}

\section{Theoretical background}
\subsection{Velocity increments vs Wavelet Transform (WT) of velocity gradients}
The classical theories of Kolmogorov \cite{K41,[K62]} are based on the scaling properties of the velocity increment, defined as $\delta_\ell \textbf{u}={\bf u}({\bf x+r})-{\bf u}({\bf x})$, where $\ell=\vert {\bf r}\vert$ is the distance over which the increment is taken.
As pointed out by \cite{[M91]}, a more natural tool to characterize the  local scaling properties of the velocity field is the wavelet transform of the tensor $\partial_j u_i $, defined as:
 \begin{equation}
G_{ij}({\textbf{x}},{\ell})=\int \mathrm{d} {\textbf{r}}\nabla_j \Phi_\ell \left(\bf{r}\right) 
u_i(\bf{x}+\bf{r}),
 \label{wavelets}
 \end{equation}
where $\Phi_\ell(\textbf{x})=\ell^{-3}\Phi(\textbf{x}/\ell)$ is a smooth function, non-negative with unit integral. 
In what follows, we choose a Gaussian function
$\Phi(\textbf{x})=\exp(-\|\textbf{x}\|^2/2)/(2\pi)^{\frac32}$ such that $ \int \Phi(\textbf{r})\mathrm{d}\textbf{r}=1$.  
We then compute the wavelet velocity increments as
\begin{equation}
\delta W(\textbf{u})(\textbf{x},\ell)= \ell \max_{ij} \vert G_{ij}(\textbf{x},\ell)\vert.
\label{waveincre}
\end{equation}
This formulation is especially well suited for the analysis of the experimental velocity field, as it naturally allows to
average out the noise. It has been verified that the wavelet based approach yields the same values for the
scaling exponents as those computed from the velocity increments \cite{Debue18}.

\subsection{K41 and K62 universality}
\label{sec:K41}
In the first theory of Kolmogorov \cite{K41}, the turbulence properties depend only on two parameters:  the mean energy dissipation per unit mass $\epsilon$ and the viscosity $\nu$ . The only velocity and length unit that one can build using these quantities are the Kolmogorov length $\eta=(\nu^3/\epsilon)^{1/4}$ and velocity $u_{\mathrm{K}}=(\epsilon \nu)^{1/4}$. The structure functions are then self-similar in the inertial range $\eta\ll \ell\ll L_0$, where $L_0$ is the integral scale, and follow the universal scalings:
\begin{equation}
	S_{p}(\ell)\equiv \langle \vert\delta_\ell u\vert^{p} \rangle\sim u_{\mathrm{K}} ^{p} \left(\frac{\ell}{\eta}\right)^{p/3},
	\label{K41Universality}
\end{equation}
which can also be recast into:
\begin{equation}
\tilde S_p\equiv \frac{S_p}{S_3^{p/3}}=C_p,
	\label{K41UniversalityFin}
\end{equation}
where $C_p$ is a (non universal) constant.

This scaling is typical of a global scale symmetry solutions, and was criticized by Landau, who considered it incompatible with observed large fluctuations of the local energy dissipation. Kolmogorov then built a second theory (K62), in which fluctuations of energy dissipation were assumed to follow a log-normal statistics, and taken into account via an intermittency exponent $\mu$ and a  new length scale $L$, thereby breaking the global scale invariance. The resulting velocity structure functions then follow the new scaling:
\begin{equation}
	S_{p}(\ell)\sim (\epsilon \ell) ^{p/3} \left(\frac{\ell}{L}\right)^{\mu p (3-p)},
	\label{K62Scaling}
\end{equation}

which implies a new kind of universality involving the relative structure functions $\tilde S_p$ as:
\begin{equation}
	\tilde S_p\equiv \frac{S_p}{S_3^{p/3}}\sim A_p\left(\frac{\ell}{L}\right)^{\tau(p)},
	\label{K62Universality1}
\end{equation}
	where $\tau(p)=\mu p(3-p)$ and $A_p$ is a constant. Such a formulation already predicts an interesting universality, if $L=L_0$, as we should have:
\begin{equation}
	\left(\frac{L_0}{\eta}\right)^{\tau(p)}\tilde S_p\sim A_p\left(\frac{\ell}{\eta}\right)^{\tau(p)}.
	\label{K62Universality1bis}	
	\end{equation}
Therefore, we should be able to collapse all structure functions, at different Reynolds number by plotting $(\frac{L_0}{\eta})^{\tau(p)}\tilde S_p$ as a function of $\frac{\ell}{\eta}$, given that  $L_0/\eta\sim \text{Re}^{3/4}$. There is however no clear prediction about the value of 
	$L$ and we show in the data analysis section that $L$ differs from $L_0$. \
	
	The relation \eqref{K62Universality1bis} shows that $\log\left(\left(\frac{L_0}{\eta}\right)^{\tau(p)}\tilde S_p\right)$ is a linear function of $\log(\frac{\ell}{\eta})$. In principle, such universal scaling is not valid outside the inertial range, i.e. for example when $\ell<\eta$. To be more general than previously thought, it can however be shown using the multifractal formalism as first shown by \cite{Frisch_1991}.

\subsection{Multifractal and fluctuating dissipation length}
For the multifractal (MFR) model, it is assumed that  the turbulence is locally self-similar, so that there exists   a scalar field $h(\textbf{x}, \ell,t)$, such that
\begin{equation}
h\left(\textbf{x}, t, \ell\right) = \frac{\log\left(\vert\delta_{\ell} \textbf{u}(\textbf{x}, t)\vert/u_0\right)}{\log(\ell/L)},
\label{LocalHolderatscaler}
\end{equation}
for a range of scales in a suitable "inertial range"  $\eta_h\ll \ell\ll L$, where $L$ is a characteristic integral-length-scale, $\eta_h$ a cut-off length scale, and $u_0$ a characteristic large-scale velocity. This scale is a generalization of the Kolmogorov scale, and is defined as the scale where the local Reynolds number $\ell |\delta_{\ell} \bm u|/\nu$ is equal to $1$. Writing  $\delta_{\ell} u=|\delta_{\ell} \textbf{u}|=u_0(\ell/L)^h$ leads to  the expression of $\eta_h$ as a function of the global Reynolds number $\text{Re}=u_0 L/\nu$ as $\eta_h\sim L \text{Re}^{-1/(1+h)}$. 
This scale  thus appears as a fluctuating cut-off which depends on the scaling exponent and therefore on $\bm x$. This is the generalization of the Kolmogorov scale $\eta \sim \nu^{3/4}\equiv \eta_{\frac13}$, and was first proposed in \cite{[P87]}. Below $\eta_h$, the velocity field becomes laminar, and $|\delta_\ell \textbf{u}|\propto \ell$.\
When the velocity field is turbulent, $h\equiv \log(|\delta_{\ell} \textbf{u}|/u_0)/\log(\ell/L)$ varies stochastically as a function of space and time. Also, if the turbulence is statistically homogeneous, stationary and isotropic, $h$ only depends on $\ell$, the scale magnitude. Therefore, formally,  $h$ can be regarded as a continuous stochastic process labeled by $\log(\ell/L)$. By Kramer's theorem \citep{Touchette09}, one sees that as in the limit $\ell\to 0$, $\log(L/\ell)\to \infty$, we have
\begin{equation}
{\mathbb{P}} \left[ \log(\delta_\ell u /u_0)= h\log(\ell/L)\right] \sim e^{\log(\ell/L)C(h)}=\left(\frac{ \ell}{L}\right)^{C(h)},
\label{LargedeviationC}
\end{equation}
where $C(h)$  is the rate function of $h$, also called multifractal spectrum. Formally, $C(h)$ can be interpreted as  the co-dimension of the set where the local H\"older exponent at scale $\ell$ is equal to $h$.
Using G\"artner-Elis theorem \citep{Touchette09}, one can connect $C$ and the velocity structure functions as:
\begin{equation}
S_p(\ell)=\langle(\delta_\ell u)^p\rangle= \int\limits_{h_{\min}}^{h_{\max}} u_0^p\left(\frac{\ell}{L}\right)^{ph+C(h)} \mathrm{d}h.
\label{LargedeviationLun}
\end{equation}
To proceed further and make connection with previous section, we set $\epsilon=u_0^3/L$ so that  $S_p(\ell)$ can now be written:
\begin{equation}
S_p(\ell)=(\epsilon \ell)^{p/3} \int\limits_{h_{\min}}^{h_{\max}}\Big(\frac{\ell}{L}\Big)^{p(h-1/3)+C(h)} \mathrm{d}h \sim (\epsilon \ell)^{p/3} \Big(\frac{\ell}{L}\Big)^{\tau(p)}.
\label{LargedeviationL}
\end{equation}
This shows that $\tau(p)$ is the Legendre transform of the rate function $C(h+1/3)$, i.e.  $\tau(p)=\min_h(p(h-1/3) +C(h))$, and equivalently, that $C(h)$ is the Legendre transform of $\tau(p)$.  Because of this,  it is necessarily convex. The set of points where it satisfies $C(h)\le d$, represents the set of admissible or observable $h$, is therefore necessarily an interval, bounded by $-1\le h_{\min}$ and $h_{\max}\le 1$.\

As noted by \cite{Frisch_1991}, the scaling exponent $\zeta(p)=p/3+\tau(p)$ defined via Equation \eqref{LargedeviationL} is only constant in a range of scale where $\ell> \eta_h$ for any $h\in [h_{\min}, h_{\max}]$. For small enough $\ell$, this condition is not met anymore, since as soon as $\ell<\eta_h$, all velocity fields corresponding to $h$ are "regularized", and do not contribute anymore to intermittency since they scale like $\ell$. 
This results in a slow dependence of $\zeta(p)$ with respect to the scale, which is obtained  via the corrected formula:
\begin{equation}
S_p=(\epsilon \ell)^{p/3} \int\limits_{\eta_h\le \ell}\Big( \frac{\ell}{L}\Big)^{p(h-1/3)+C(h)} \mathrm{d}h \sim (\epsilon \ell)^{p/3}\Big(\frac{\ell}{L}\Big)^{\tau(p,\ell)}.
\label{LargedeviationCorrected}
\end{equation}
To understand the nature of the correction, we can compute the value of $h$ such that $\ell=\eta(h)$. It is simply: $h(\ell)=-1+\log(\text{Re})/\log(L/\eta)$. We note $\theta=\log(L/\ell)/\log(\text{Re})$. We can now rewrite equation \eqref{LargedeviationCorrected} as:
\begin{equation}
\tilde S_p\equiv \frac{S_p}{S_3^{p/3}}= \int\limits_{-1+1/\theta}^{h_{\max}}\Big( \frac{\ell}{L}\Big)^{p(h-1/3)+C(h)} \mathrm{d}h \sim \exp\left(-\theta {\tau(p,\theta)\log(\text{Re})} \right),
\label{LargedeviationCorrecteddeux}
\end{equation}
where $\tau(p,\theta)=\tau_p$ when $\theta\le 1/(1+h_{\max})$ and $\tau(p,\theta)=p(\theta-1/3) +C(-1+1/\theta)$ when $1/(1+h_{\max})\le \theta\le 1/(1+h_{\min})$. As discussed by \cite{Frisch_1991}, this implies a new form of universality that extends beyond the inertial range, into the so-called extended dissipative range, as;
\begin{equation}
	\frac{\log({\tilde S_{p}})}{\log(\text{Re})}=-\tau(p,\theta)\theta, \quad 	\theta=\log(L/\ell)/\log(\text{Re}).
\label{eq:Sp_univ}
\end{equation}
If the scale $L$ is constant and equal to $L_0$, the integral scale, then we have $\text{Re}=(L_0/\eta)^{4/3}$ and the multifractal universality implies that  $\log(\tilde S_{p})/\log(L_0/\eta)$ is a  function of $\log(\ell/\eta)/\log(L_0/\eta)$. When the function is linear, we thus recover the K62 universality. The multifractal universaility is thus a {\sl generalization} of the K62 universality.

This form of universality is however not easy to test, as the scale $L$ is not known a priori, and may still depend on $\text{Re}$. In what follows, we  demonstrate a new form of universality, that allows more freedom upon $L$ and encompass both K62 and multifractal universality.

\subsection{General universality}
Using the hypothesis that turbulence maximizes some energy transfer in the scale space, Castaing \cite{Castaing89} suggested a new form of universality for the structure functions, that reads:
\begin{equation}
	\gamma(\text{Re})\log \left(\frac{S_p}{A_{p}u_{\mathrm{K}}^p}\right)= G\left(p,\gamma(\text{Re})\log(\ell K_0/\eta) \right),
	\label{Castaing}
\end{equation}
where $A_{p}$ and $K_0$ are universal constant and $\beta$ and $G$ are general functions, $F$ being  linear in the inertial range, $G(p,x)\sim \tau(p) x$.
The validity of this universal scaling was checked by Gagne and Castaing \cite{Castaing93} on data obtained from the velocity fields measured in a jet using hot wire anenometry. They found good collapse of the structure functions at different Taylor Reynolds $R_\lambda$, provided $\gamma(\text{Re})$ is constant at low Reynolds numbers and  follows  a law of the type: $\gamma(\text{Re})\sim \gamma_0/\log(R_\lambda/R_*)$, where $R_*$ is a constant, whenever $R_\lambda>400$. Since we have $R_\lambda\sim \text{Re}^{1/2}$ and $(L_0/\eta)\sim \text{Re}^{3/4}$, we can rewrite equation \eqref{Castaing} as:
\begin{equation}
	\beta(\text{Re})\left(\frac{\log( \tilde S_p/S_{0p})}{\log(L_0/\eta)}\right)= G\left(p,\beta(\text{Re})\frac{\log(\ell/\eta)}{\log(L_0/\eta)} \right),
	\label{Castaingnew}
\end{equation}
where $S_{0p}$ are some constants and $\beta$ and $F$ are general functions. Comparing with the K62 or MFR universality formulae \eqref{K62Universality1bis} or \eqref{eq:Sp_univ}, we see that formula \eqref{Castaingnew} is a generalization of these two universality with $L=L_0$. It allows however more flexibility than K62 or MFR universality through the function $\beta(\text{Re})$, that is a new fitting function. We test these predictions in Section \ref{Sec:check} and provide a physical interpretation of \eqref{Castaingnew}  in Section \ref{Sec:thermo}.

\section{Check of universality using data analysis}
\label{Sec:check}
The various universality are tested using the velocity structure functions based on the wavelet velocity increments Eq. \eqref{waveincre}, in order to minimize the noise in the experimental data. We define:
\begin{equation}
S_p=\langle\vert \delta W(\textbf{u})(\textbf{x},\ell)\vert^p\rangle.
\label{wavestrucfonc}
\end{equation}
We then apply this formula to both experimental data (Table \ref{tab:exp_parameter}) and numerical data (Table \ref{tab:DNS_parameter}), to get wavelet velocity structure functions at various scales and Reynolds numbers.

\subsection{Check of K41 universality}
\begin{figure}
	\includegraphics[width=0.49\linewidth]{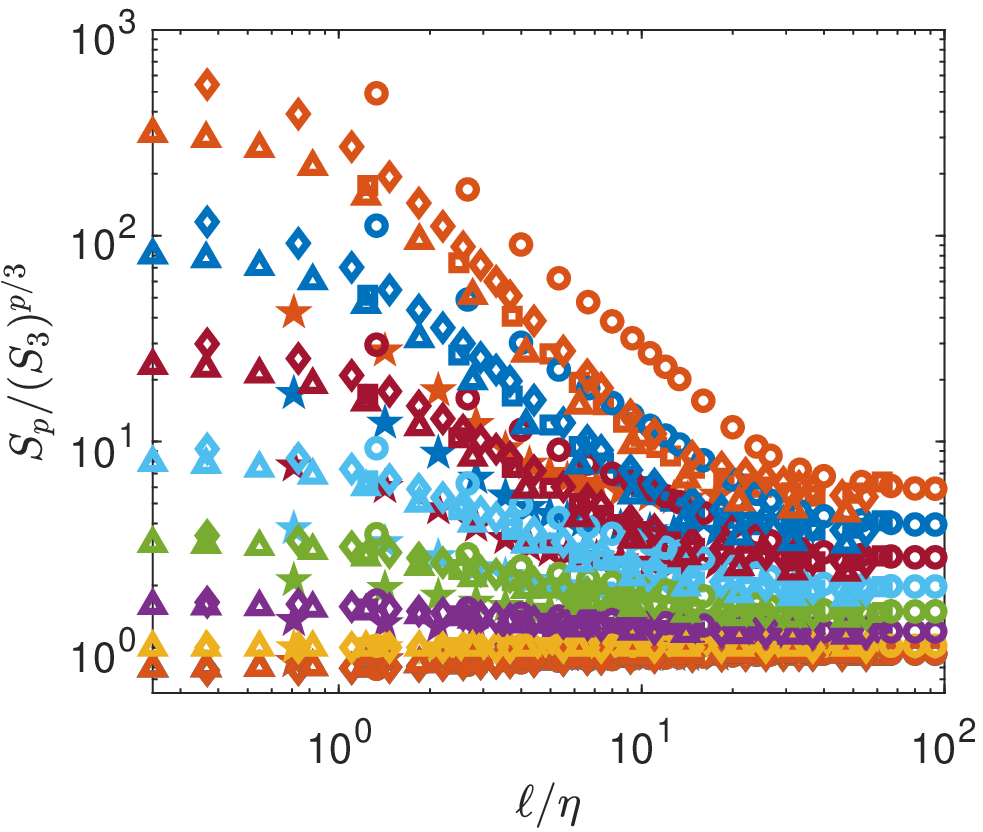}
	\put(-40,150){\bf(a)}
	\includegraphics[width=0.49\linewidth]{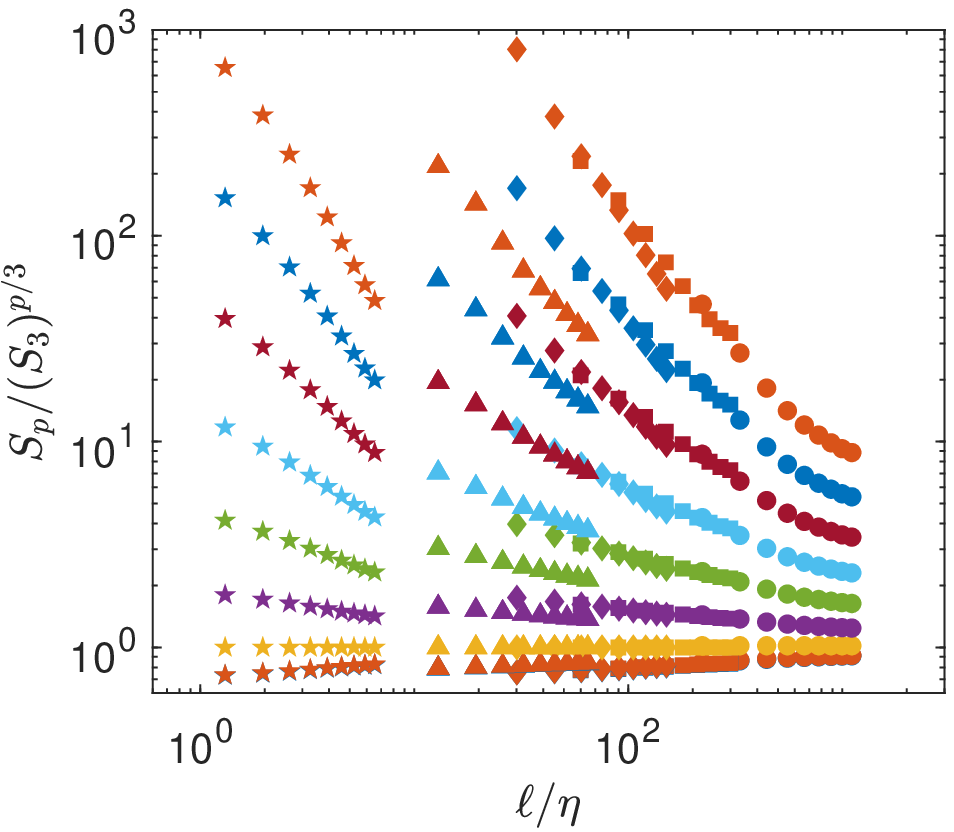}
	\put(-40,150){\bf(b)}
\caption{Test of K41 universality Eq. \eqref{K41UniversalityFin}. a) Numerical data b) Experimental data. The structure functions have been shifted by arbitrary factors for clarity and are coded by color: $p=1$: blue symbols; $p=2$: orange symbols; $p=3$: yellow symbols; $p=4$: magenta symbols; $p= 5$: green symbols; $p=6$: light blue symbols; $p=7$: red symbols; $p= 8$: blue symbols; $p=9$: orange symbols. For K41 universality to hold, all the function should be constant, for a given $p$.}
\label{Fig:checkK41}
\end{figure}

The K41 universality \eqref{K41Universality} can be checked by plotting:
\begin{equation}
\log\left(\frac{S_p}{u_{\mathrm{K}}^p}\right)=F\left(\log\left(\frac{\ell}{\eta}\right)\right).
\label{checkK41}
\end{equation}
This is shown in figure \ref{Fig:checkK41} for both experimental and numerical data. Obviously, the data do not collapse on a universal curve, meaning that K41 universality does not hold. This is well known, and is connected to intermittency effects \cite{FB}.

\subsection{Check of K62 universality}
The K62 universality \eqref{K62Universality1bis} can be checked by plotting:
\begin{equation}
\log\left[\left(\frac{L_0}{\eta}\right)^{\tau(p)} \tilde S_p\right]=F\left(\log\left(\frac{\ell}{\eta}\right)\right).
\label{check62}
\end{equation}

\begin{figure}
	\includegraphics[width=0.49\linewidth]{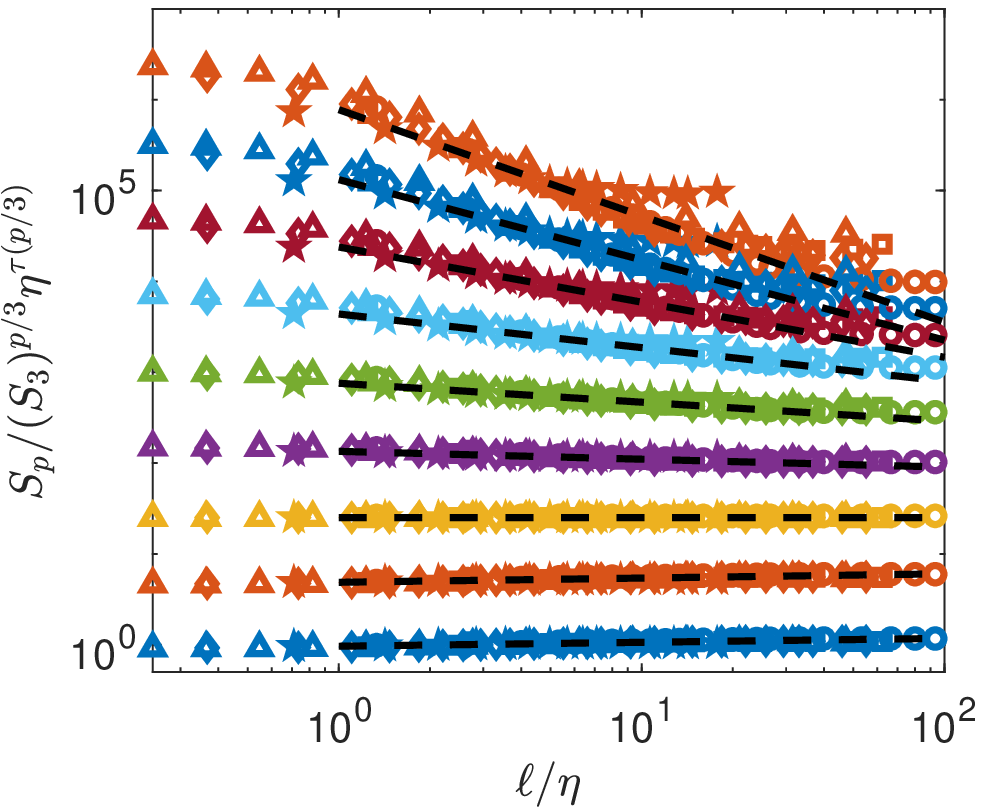}
	\put(-40,150){\bf(a)}
	\includegraphics[width=0.49\linewidth]{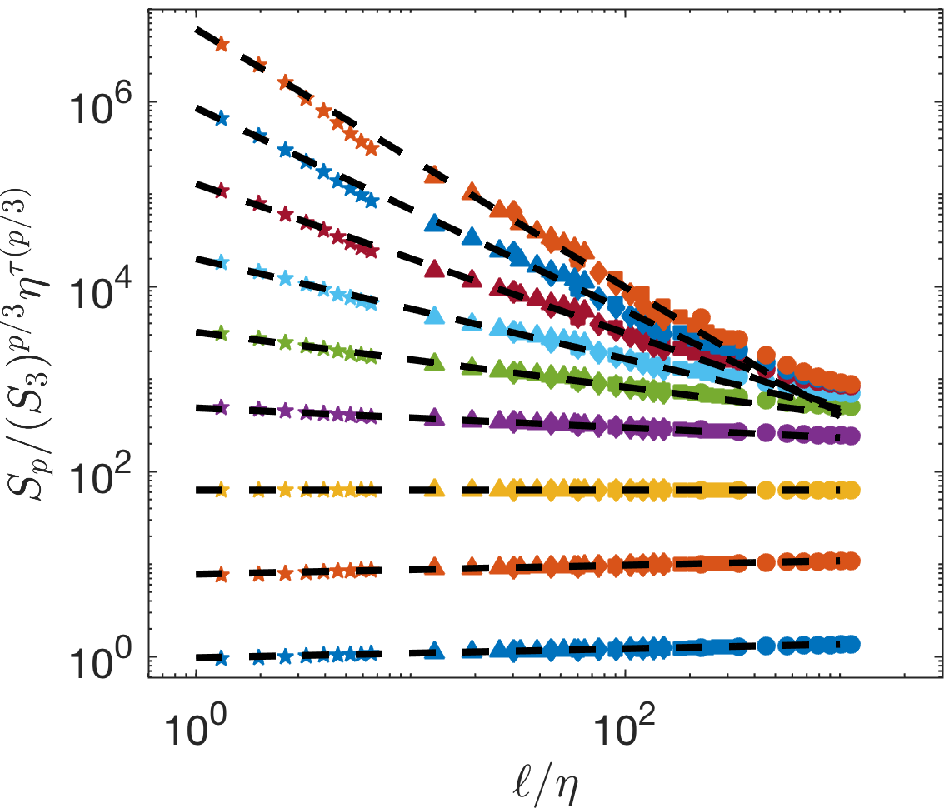}
	\put(-40,150){\bf(b)}
\caption{Test of K62 universality Eq. \eqref{K62Universality1bis}. a) Numerical data b) Experimental data. The structure functions have been shifted by arbitrary factors for clarity and are coded by color: $p=1$: blue symbols; $p=2$: orange symbols; $p=3$: yellow symbols; $p=4$: magenta symbols; $p= 5$: green symbols; $p=6$: light blue symbols; $p=7$: red symbols; $p= 8$: blue symbols; $p=9$: orange symbols. The dashed lines are power laws with exponents $\tau(p)=\zeta(p)-\zeta(3) p/3$, with $\zeta(p)$ shown in figure \ref{Fig:scalings}-a.}
\label{Fig:checkK62}
\end{figure}

The  collapse depends directly on $\tau(p)$, the intermittency exponents. Obtaining the best collapse of all curves is in fact a way to fit the best scaling exponents $\tau(p)$. We thus implemented a minimization algorithm that provides the values of $\tau(p)$ that minimized the distance between the curve and the line of slope $\tau(p)$. The values of $\tau(p)$ are reported in Table \ref{tab:zeta_p}. The best collapse  is shown on Figure \ref{Fig:checkK62}-a for the DNS, and Figure \ref{Fig:checkK62}-b for the experiment.
The collapse is  better for experiments than for the DNS. However, in both cases, there are significant differences in between points at different $R_\lambda$, at larger scales, showing that universality is not yet reached.

\begin{figure}
	\includegraphics[width=0.49\linewidth]{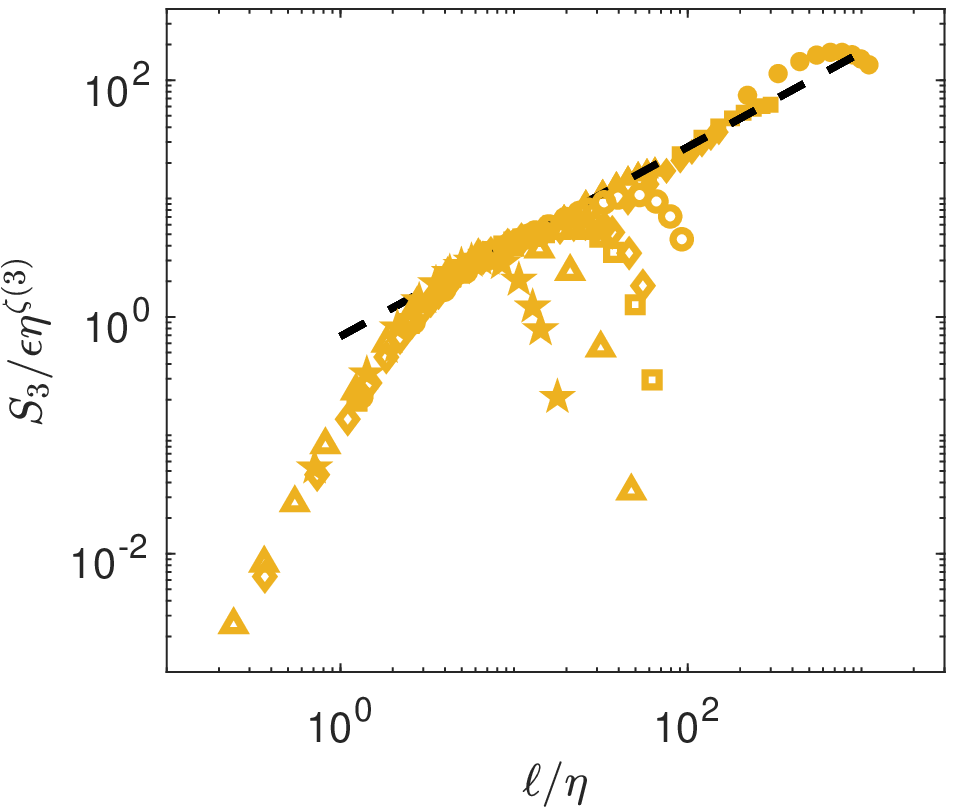}
	\put(-40,40){\bf(a)}
	\put(-100,140){$\ell^{0.8}$}
	\includegraphics[width=0.49\linewidth]{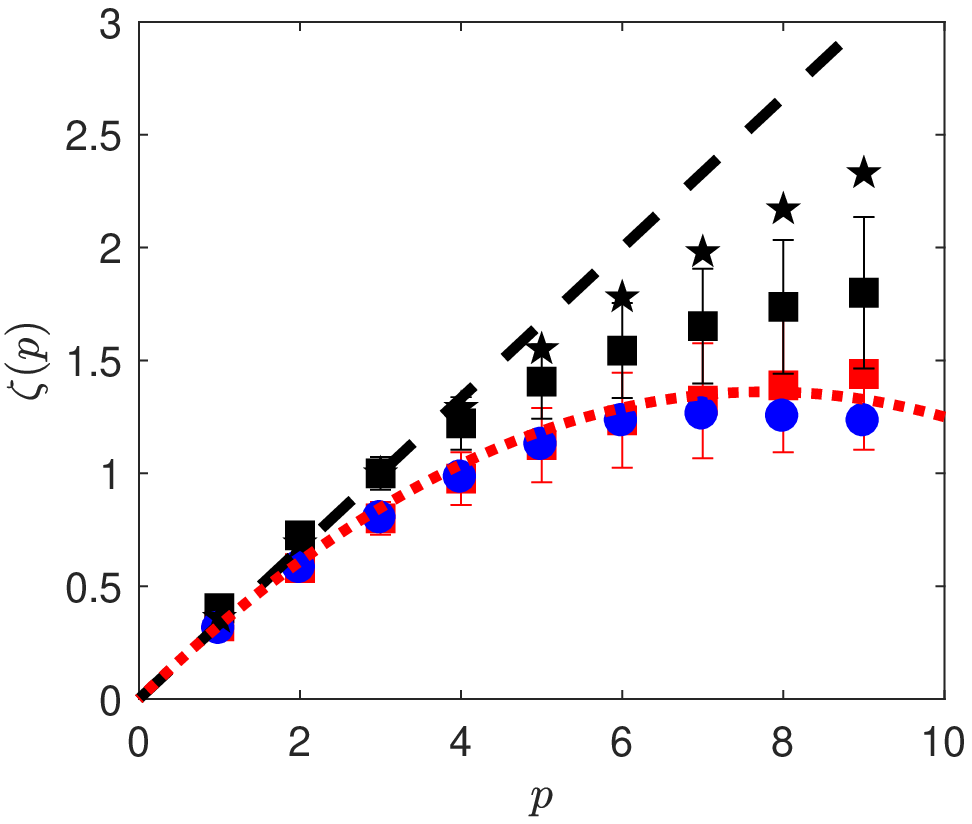}
	\put(-40,40){\bf(b)}
\caption{ . a) Determination of $\zeta(3)$ by best collapse using both DNS (open symbols)  and experiments (filled symbols). The black dashed line is $\ell^{0.8}$.  b)  Scaling exponents $\zeta(p)$ of the wavelet structure functions of $\delta W$ as a function of the order, from Table \ref{tab:zeta_p}, for DNS (blue circle) and experiments (red square) .    The red dotted line is the function $\min_{h}(hp+C(h))$ with $C(h)$ given by $C(h) = (h-a)^2/2b$,
with  $a=0.35$ and $b=0.045$. The black stars correspond to $\zeta_{\text{SAW}}(p)/\zeta_{\text{SAW}}(3)$ (see Table \ref{tab:zeta_p}), while the black triangle correspond to $\zeta_{\text{EXP}}(p)/\zeta_{\text{EXP}}(3)$. }
\label{Fig:scalings}
\label{fig:zeta_p}
\end{figure}

\subsection{Check of General Universality}
\begin{figure}
	
	\includegraphics[width=0.49\linewidth]{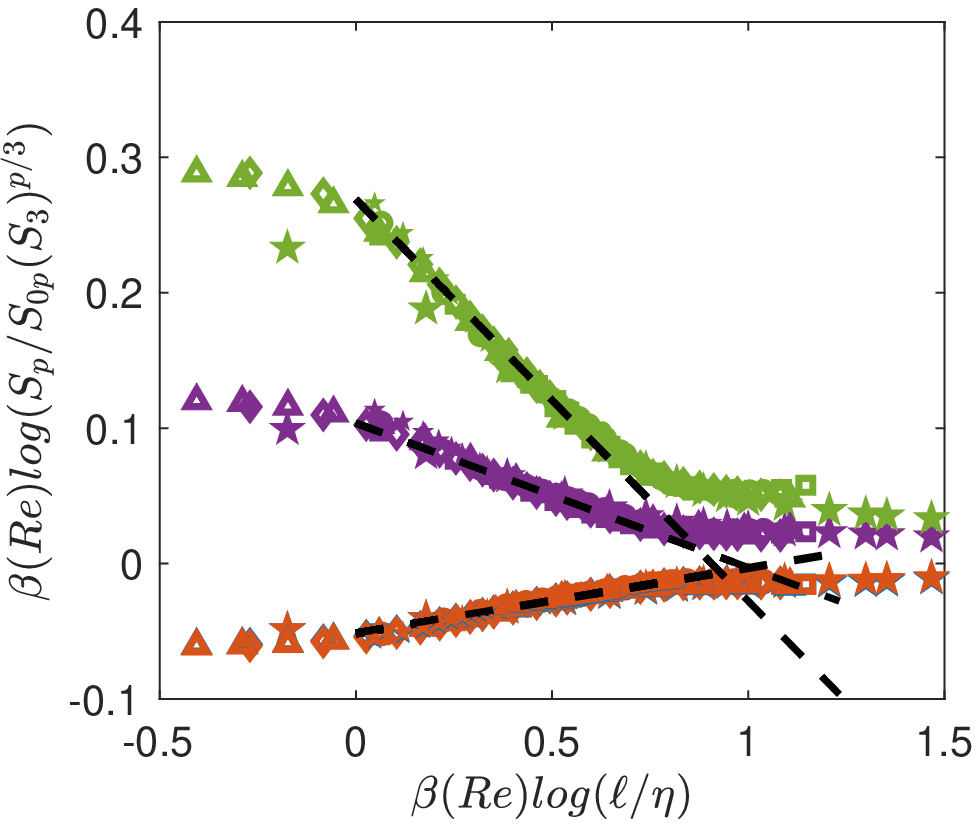}
	\put(-130,30){\bf(a)}
	\includegraphics[width=0.49\linewidth]{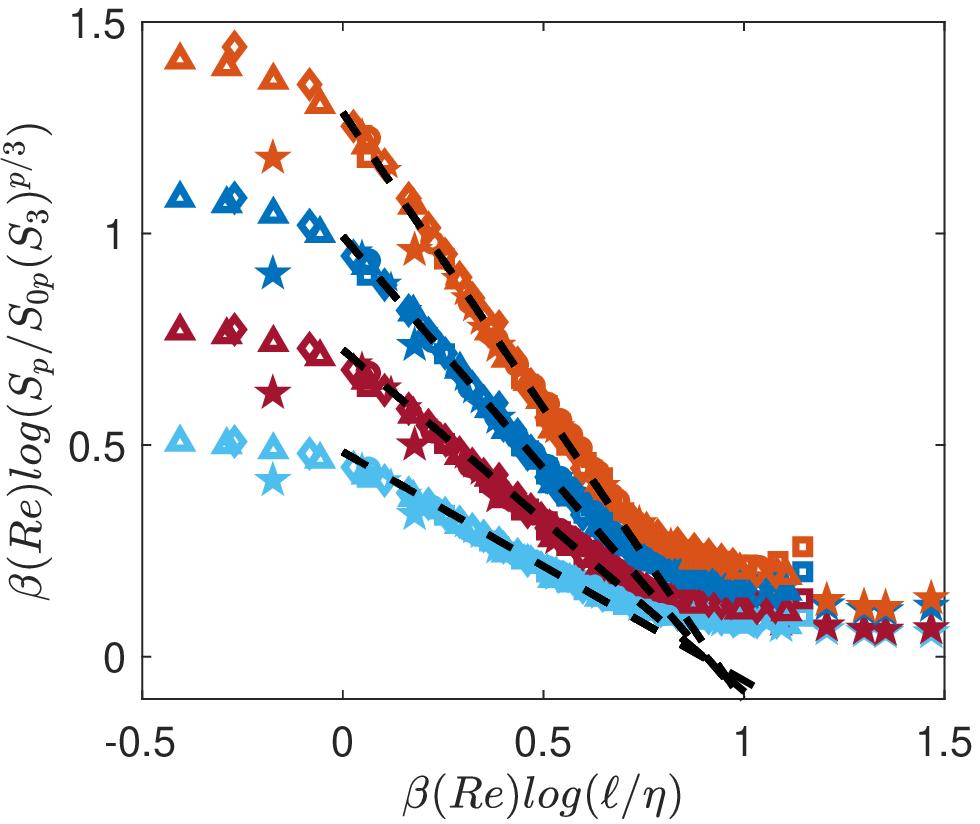}
	\put(-130,30){\bf(b)}
\caption{Test of general universality equation \eqref{checkGuniversal} using both DNS (open symbols)  and experiments (filled symbols). The  functions  are coded by color. a) $p=1$: blue symbols; $p=2$: orange symbols; $p=4$: magenta symbols; $p= 5$: green symbols;  b) $p=6$: light blue symbols; $p=7$: red symbols; $p= 8$: blue symbols; $p=9$: orange symbols. The  functions have been shifted by arbitrary factors for clarity. The dashed lines are power laws with exponents $\tau(p)=\zeta(p)-\zeta(3) p/3$, with $\zeta(p)$ shown in figure \ref{Fig:scalings}-a.}
\label{Fig:checkGUniv}
\end{figure}

We can now check the most general universality, by plotting:
\begin{equation}
\beta(\text{Re})\left(\frac{\log( \tilde S_p/S_{0p})}{\log(L_0/\eta)}\right)= F\left(p,\beta(\text{Re})\frac{\log(\ell/\eta)}{\log(L_0/\eta)} \right),
\label{checkGuniversal}
\end{equation}
In this case, best collapse is obtained by fitting two families of parameters: $S_{0p}$, $\beta(\text{Re})$ that were obtained through a procedure of minimization. We take the DNS at $R_\lambda=138$ as the reference case, and find for both DNS and experiments,  the values of $\beta(\text{Re})$ and $S_{0p}$ that best collapse the curves. The corresponding collapses are provided in figure \ref{Fig:checkGUniv}. The collapse is good for any value of $\text{Re}$, except for the DNS at the lowest Reynolds number, which does not collapse in the far dissipative range.

\subsection{Function $\beta(\text{Re})$}
\begin{figure}
	\includegraphics[width=0.49\linewidth]{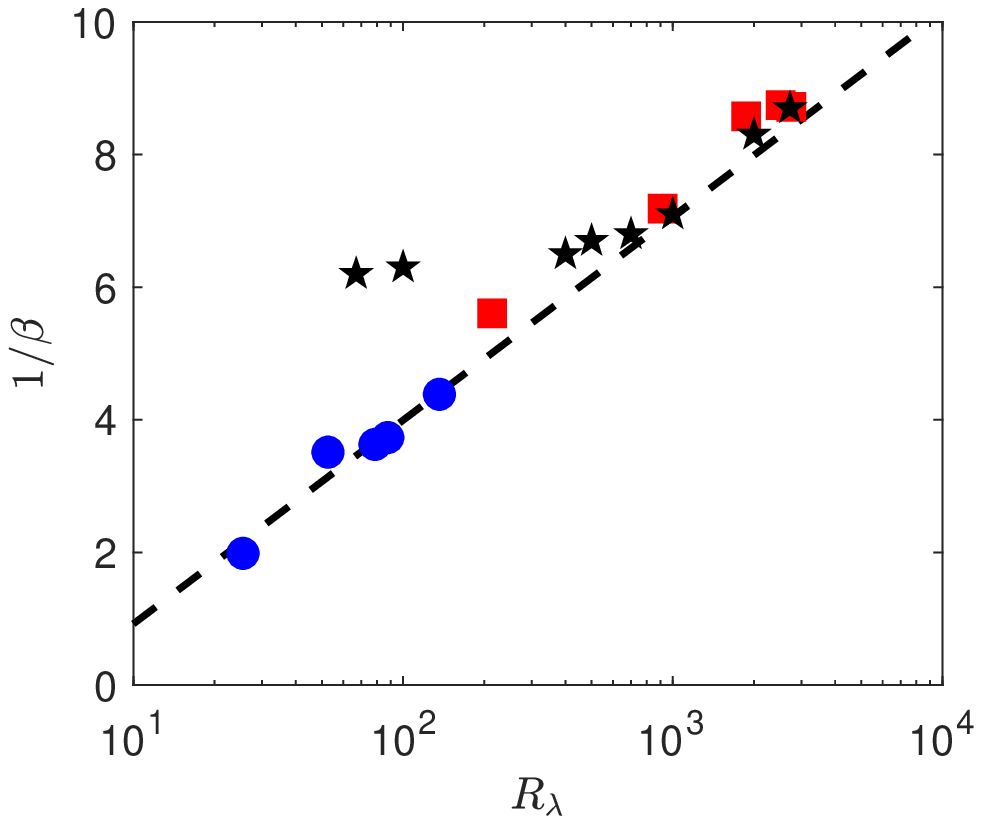}
	\includegraphics[width=0.49\linewidth]{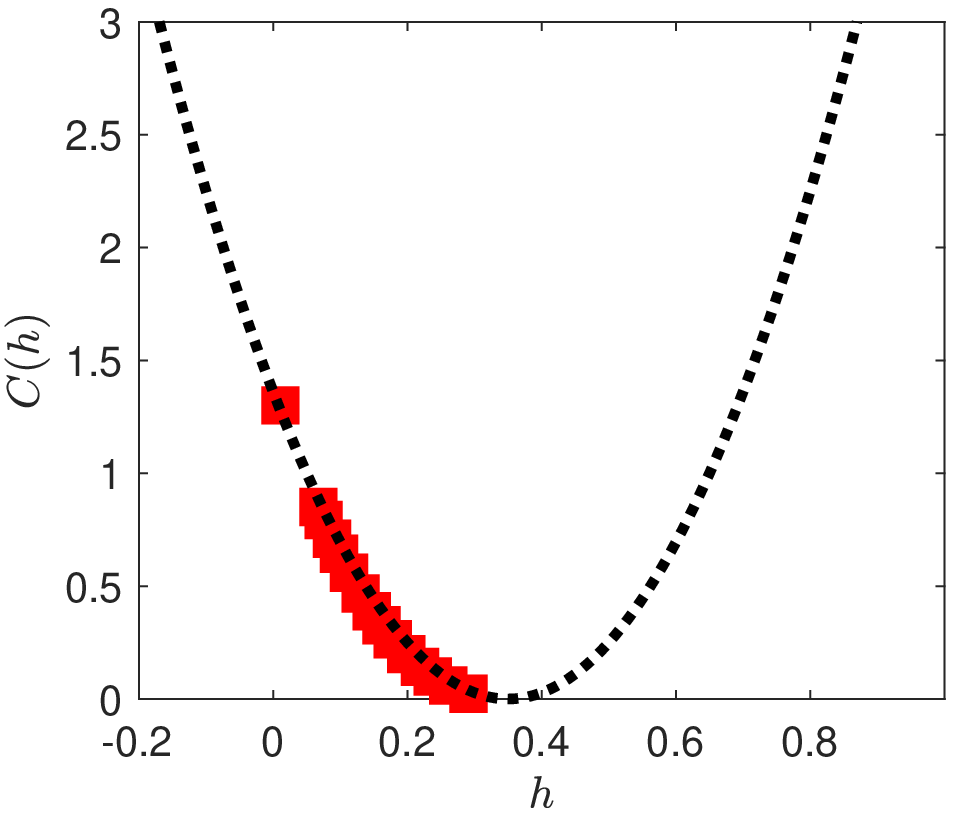}
	\caption{ a)  Variation of $1/\beta(\text{Re})$ versus $\log(R_\lambda)$ in experiments (red square) and DNS (blue circle) when using the DNS at $R_{\lambda} = 138$ as the reference case. We plotted in black the values found by Gagne and Castaing in \cite{Castaing93} shifted by an arbitrary factor to coincide the values at large Reynolds. The black dashed line is $(4/3)\log(R_\lambda/5)$. b) Multifractal spectrum $C(h)$ for the experiments. The spectrum has been obtained by taking inverse Legendre transform of the scaling exponents $\zeta(p)$ shown in figure \ref{fig:zeta_p}-b.  The dotted line is a parabolic fit $C(h)=(h-a)^2/2b$ with $a=0.35$ and $b=0.045$.}
	\label{Fig:beta}
\end{figure}

 Motivated by earlier findings by \cite{Castaing93}, we plot in figure \ref{Fig:beta} the value $1/\beta$ as a function of $R_{\lambda}$.

Our results are compatible with  $1/\beta\sim \beta_0/\log(R_{\lambda})$, with $\beta_0\sim 4/3$ over the whole range of Reynolds number.  For comparison, we provide also on figure \ref{Fig:beta} the values found by Gagne and Castaing  \cite{Castaing93} in  jet of liquid Helium, shifted by an arbitrary factor to make our values coincide with them at large Reynolds number. This shift is motivated by the fact that $\beta(\text{Re})$ is determined up to a constant, depending upon the amplitude of the structure functions used as reference. At large Reynolds, our values are compatible with theirs. At low Reynolds, however, we do not observe the saturation of $1/\beta$ that is observed in the jet experiment of \cite{Castaing93}.
An interpretation of the meaning of $\beta(\text{Re})$ will be provided in Section \ref{Sec:thermo}.

\subsection{Scaling exponents}
Our Collapse method enables us to obtain the scaling exponents of the structure functions $\zeta(p)$ by the following two methods:
\par i) Using the K62 universality, we get $\tau(p)$, and then $\zeta(p)=\zeta(3)p/3+\tau(p)$. These estimates still depend on the value of $\zeta(3)$, which is not provided by the K62 universality plot.
To obtain it, we use a minimization procedure on both experimental $\log(S_3/u_{\mathrm{K}}^3)$ from the one hand, and the numerical $\log(S_3/u_{\mathrm{K}}^3)$ on the other hand (see figure \ref{Fig:scalings}-a), to compute $\zeta(3)$ as the value that minimizes the distance between the curve and a straight line of slope $\zeta(3)$. The values so obtained are reported in Table \ref{tab:zeta_p}, and have been used to compute  $\zeta(p)$ from  $\tau(p)$.
\par ii) Using the general universality, we may also get $\tau_{p,\text{univ}}$ by a linear regression on the collapse curve. Note that since the data are collapsed, this provides a very good estimates of this quantity, with the lowest possible noise. In practice, we observe no significant differences with the two estimates; therefore, we only report the values obtained by following the first method.

The corresponding values are plotted in figure \ref{fig:zeta_p} and summarized in Table \ref{tab:zeta_p}. Note that for both DNS and experiments, the value of $\zeta(3)$ is different from $1$, which is apparently incompatible with the famous Kolmogorov $4/5$th law, that predicts $\zeta(3)=1$. This is because we use {\sl absolute} values of wavelet increments, while the Kolmogorov $4/5$th law uses signed values. We have checked that using unsigned values, we obtain a scaling that is closer to $1$, but with larger noise. Note also that when we consider the relative value $\zeta(p)/\zeta(3)$, we obtain values that are close to the values obtained \cite{saw_debue_kuzzay_daviaud_dubrulle_2018} on the same set of experimental data, using velocity increments and Extended Self-Similarity technique \cite{Benzi93}.
\begin{table}[!h]
   \centering
\begin{tabular}{@{\hspace{4mm}}l c c c c c c c cl@{\hspace{4mm}}}
   \hline
  \hline
 exponent \textbackslash\quad  order & $p=1$ & $p=2$ & $p=3$ & $p=4$ & $p=5$ & $p=6$ & $p=7$ & $p=8$ &  $p=9$  \\

  \hline  
    $\zeta_{\text{SAW}}/\zeta_{\text{SAW}}(3)$ & 0.36  & 0.69  & 1 &1.29   & 1.55 & 1.78 & 1.98 & 2.17 & 2.33\\  
    $\zeta_{\text{DNS}}$ & 0.31    &0.58    &0.80    &0.98  &1.12 &1.23 &1.26 &1.25 &1.23\\
    $\zeta_{\text{EXP}}$ & 0.32    &0.58    &0.80    &0.98    &1.12    &1.23    &1.32    &1.39    &1.44\\
    $\tau_{\text{DNS}}$ & 0.04    &0.05         &0   &-0.09   &-0.21   &-0.37  &-0.61   &-0.88   &-1.17\\
    $\tau_{\text{EXP}}$ & 0.05    &0.05         &0   &-0.09   &-0.21   &-0.36   &-0.54   &-0.74   &-0.96 \\
    \hline
    \hline
\end{tabular}
    \caption{Scaling exponents $\tau(p)$ and $\zeta(p)$ found by the collapse method based on K62 universality for experimental data (subscript EXP) or numerical data (subscript DNS).  The subscript SAW refers to the values obtained by  \cite{saw_debue_kuzzay_daviaud_dubrulle_2018} on the same set of experimental data, using velocity increments and Extended Self-Similarity technique \cite{Benzi93}. The exponents $\tau_{\text{EXP}}(p)$(red square) and $\tau_{\text{DNS}}$ (blue circle) have been computed through a least square algorithm upon $\tau(p)$, minimizing the scatter of the rescaled structure functions $\log\left[\left(\frac{L_0}{\eta}\right)^{\tau(p)} \tilde S_p)\right]$ with respect to the line $(\ell/\eta)^{\tau(p)}$. The corresponding $\zeta(p)$ were inferred using the formula $\zeta(p)=\tau(p)+\zeta(3)p/3$, where $\zeta(3)$ is computed in figure \ref{Fig:scalings}-a.}
    \label{tab:zeta_p}
\end{table}

\subsection{ Multifractal spectrum}
From the values of $\zeta(p)$, one can get the multifractal spectrum $C(h)$ by performing the inverse Legendre transform:
\begin{equation}
	C(h) = \min_{p}[ph + \zeta(p))]. 
\end{equation}
Practically, this amount to use the following formula:
\begin{equation}
	C\Big(\frac{\mathrm{d} \, \zeta(p)}{\mathrm{d} \, p}\Big|_{p^{*}}\Big) = \zeta(p^{*})-p^{*}\frac{\mathrm{d} \, \zeta(p)}{\mathrm{d} \, p} \Big|_{p^{*}}. 
\label{eq:Legendre_transform}
\end{equation}
To estimate $C$, we thus first perform a polynomial interpolation of order 4 on $\zeta(p)$, then derivate the polynom to estimate $\frac{\mathrm{d} \, \zeta(p)}{\mathrm{d} \, p}$, thus get $C$ through equation \eqref{eq:Legendre_transform}. The result is provided in figure \ref{Fig:beta}-b for both the DNS and the experiment.

The curve look like the portion of a parabola, corresponding to a log-normal statistics for the wavelet velocity increments. Specifically, fitting by the shape:
\begin{equation}
	C(h) = \frac{(h-a)^2}{2b},
	\label{fitC}
\end{equation}
we get  $a=0.35$ and $b=0.045$.  This parabolic fit also provides a good fit of the scaling exponents, as shown in figure \ref{Fig:scalings} by performing Legendre transform of $C(h)$ given by equation \eqref{fitC}.

\section{Thermodynamics and turbulence}
 \label{Sec:thermo}
\subsection{Thermodynamical analogy}
Multifractals obey  a well-known thermodynamical analogy \cite{bohr1987entropy,[M91],rinaldo1996thermodynamics} that will be useful to interpret and extend the general universality unraveled in the previous section.
Indeed, consider the quantity:
\begin{equation}
\mu_\ell=\frac{\vert\delta W_\ell\vert^3}{\langle\vert \delta W_\ell\vert^3\rangle}.
\label{definitionmu}
\end{equation}
By definition $\mu_\ell$ is positive definite and $\langle\mu_\ell\rangle=1$ for any $\ell$. It is therefore the meaning of a scale dependent measure. It then also follows a large-deviation property as:
\begin{equation}
{\mathbb{P}} \left[ \log(\mu_\ell)= E\log(\ell/\eta)\right] \sim e^{\log(\ell/\eta) S(E)},
\label{LDmeasure}
\end{equation}
where $S(E)$ is the large deviation function of $\log(\mu_\ell)$ and has the meaning of an energy while $\log(\ell/\eta)$ has the meaning of a volume, and $\log(\mu_\ell)/\log(\ell/\eta)$ is an energy density. Because of the definition of $\mu_\ell$, it is easy to see that $S$ is connected to $C$, the large deviation function of $\vert \delta W_\ell$. In fact, since in the inertial range where $\langle\vert \delta W_\ell\vert^3\rangle \sim \ell^{\zeta(3)}$, we have $S(E)=C(3h-\zeta(3))$. By definition, we also have:
\begin{equation}
\tilde S_{3p}=\frac{S_{3p}}{S_3^{p}}=\langle e^{p\log(\mu_\ell)}\rangle,
\label{defistilde}
\end{equation}
so that $Z=\tilde S_{3p}$ is the partition function associated to the variable $\log(\mu_\ell)$, at the pseudo-inverse temperature $p=1/kT$. Taking the logarithm of $Z$, we then get the free energy $F$ as:
\begin{equation}
\log(\tilde S_{3p})=F.
\label{freeenergy}
\end{equation}
By the G\"artner-Elis theorem, $F$ is the Legendre transform of the energy $S$: $F=\min_E(pE-S(E))$. The free energy a priori depends on the temperature; i.e. on $T=1/kp$, on the volume $V=\log(\ell/\eta)$ and on the number of degrees of freedom system $N$. If we identify $N=(1/\beta(\text{Re}))\log(L_0/\eta)$, we see that the general universality means:
\begin{equation}
F(T,V,N)=NF(T,\frac{V}{N},1),
\label{extensivity}
\end{equation}
i.e. can be interpreted as   {\sl extensivity } of the free energy.  

\begin{table}[!h]
\centering
\begin{tabular}{c@{\hspace{20mm}}c|@{\hspace{10mm}}c}
    &Thermodynamics & Turbulence \\
\hline
Temperature &$ k_{B}T$ & $ 1/p$\\
Energy &$E$ & $\log(\mu_\ell)$\\
Number of d.f. &$N$ & $\log(\text{Re})\equiv \log(L_0/\eta)/\beta_0$\\
Volume &$V$ & $\log(\ell/\eta)$\\
Pressure &$P$ & $\tau(p,\ell)$\\
Free energy &$F$ & $\log(\tilde S_{3p})$\\
\end{tabular}
\caption{Summary of the analogy between the multifractal formalism of turbulence and thermodynamics.}
	\label{tab:analogy}
\end{table} 
The thermodynamic analogy is thus meaningful and is summarized in Table \ref{tab:analogy}.
It can be used to derive interesting prospects.

\subsection{Multifractal pressure and phase transition}
\begin{figure}
	\centering
	\includegraphics[width=0.49\linewidth]{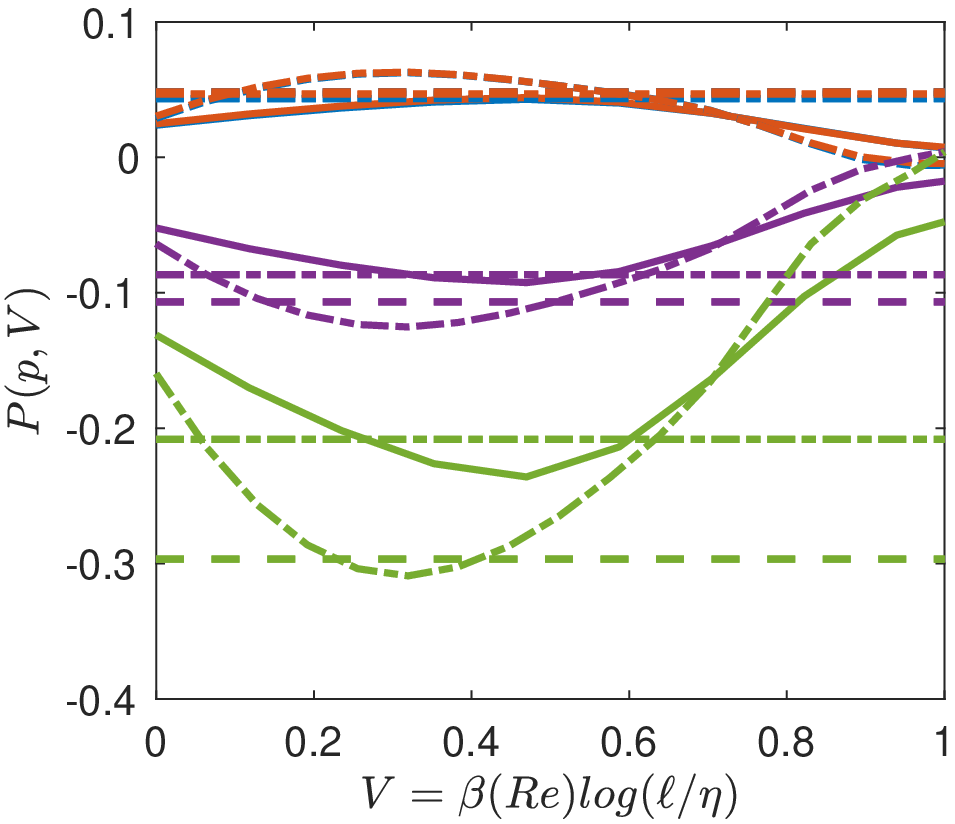}
	\includegraphics[width=0.49\linewidth]{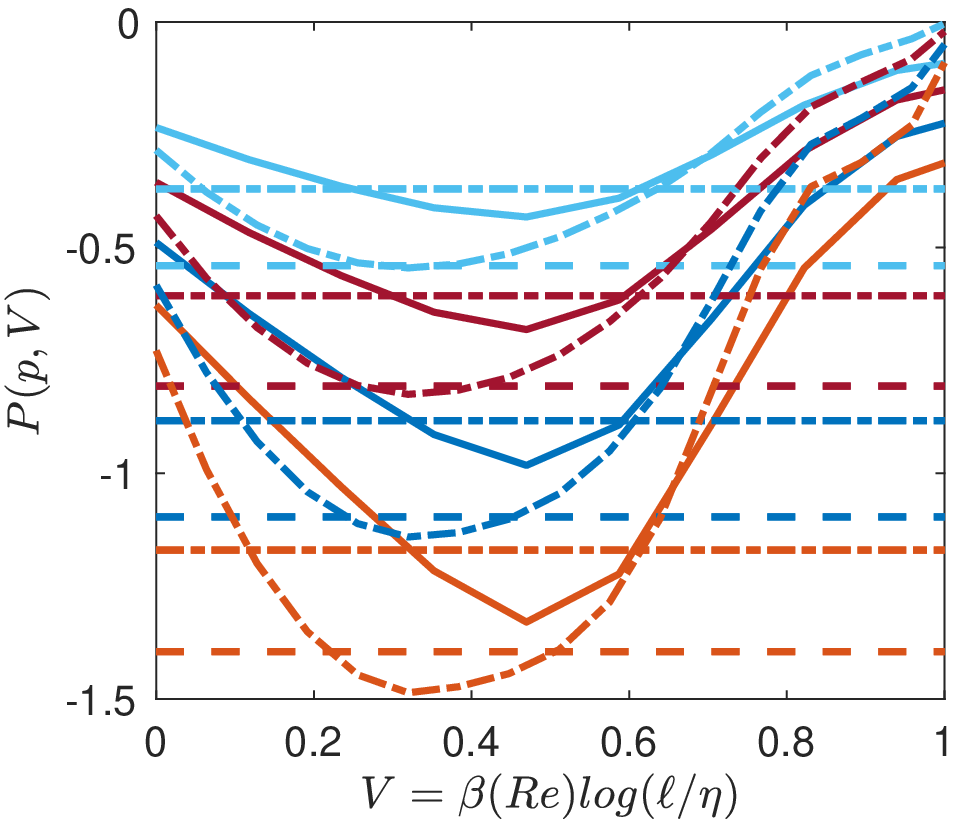}

	\caption{Multifractal equation of state of turbulence. Multifractal pressure as a function of the volume for $ \ R_{\lambda} = 25$ (line) , $\ R_{\lambda} = 56$ (dashed-dotted line). The  functions  are coded by color. a) $p=1$: blue symbols; $p=2$: orange symbols; $p=4$: magenta symbols; $p= 5$: green symbols;  b) $p=6$: light blue symbols; $p=7$: red symbols; $p= 8$: blue symbols; $p=9$: orange symbols. The colored dotted line (resp. dashed dotted line) are values corresponding to $P(p,V)=\tau_{\mathrm{EXP}}(p)$ (resp. $\tau_{\mathrm{DNS}}(p)$, that are reported in Table \ref{tab:zeta_p}.}
\label{fig:pressurelam}
\end{figure}

\begin{figure}
	\centering
	\includegraphics[width=0.49\linewidth]{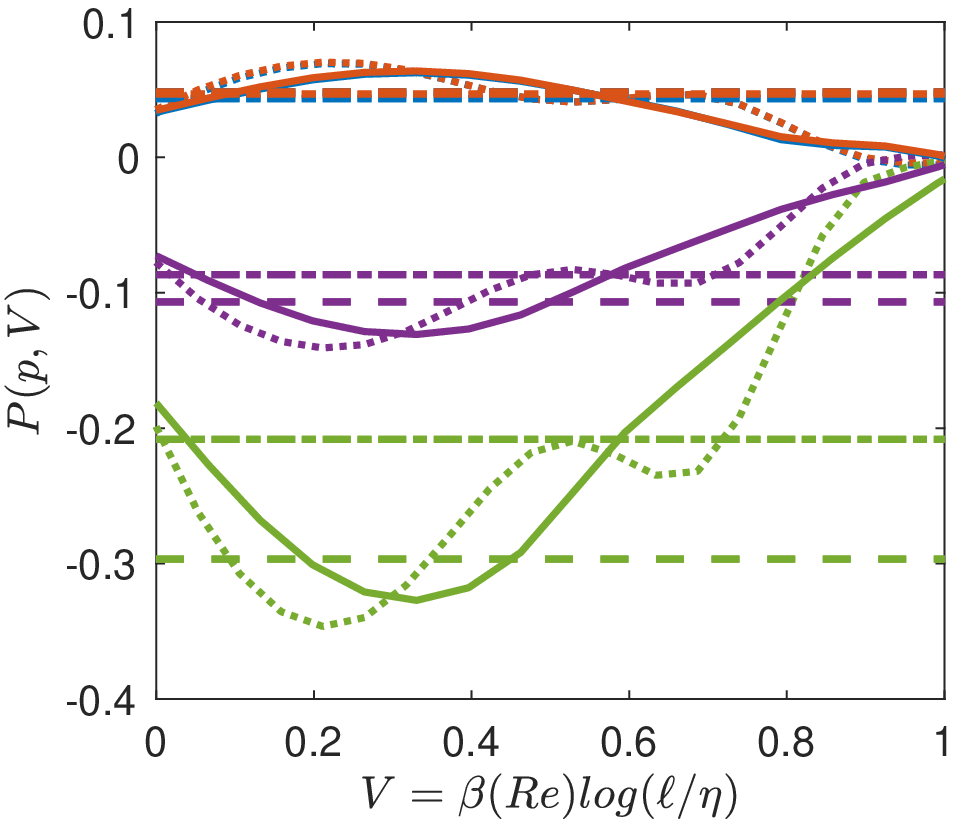}
	\includegraphics[width=0.49\linewidth]{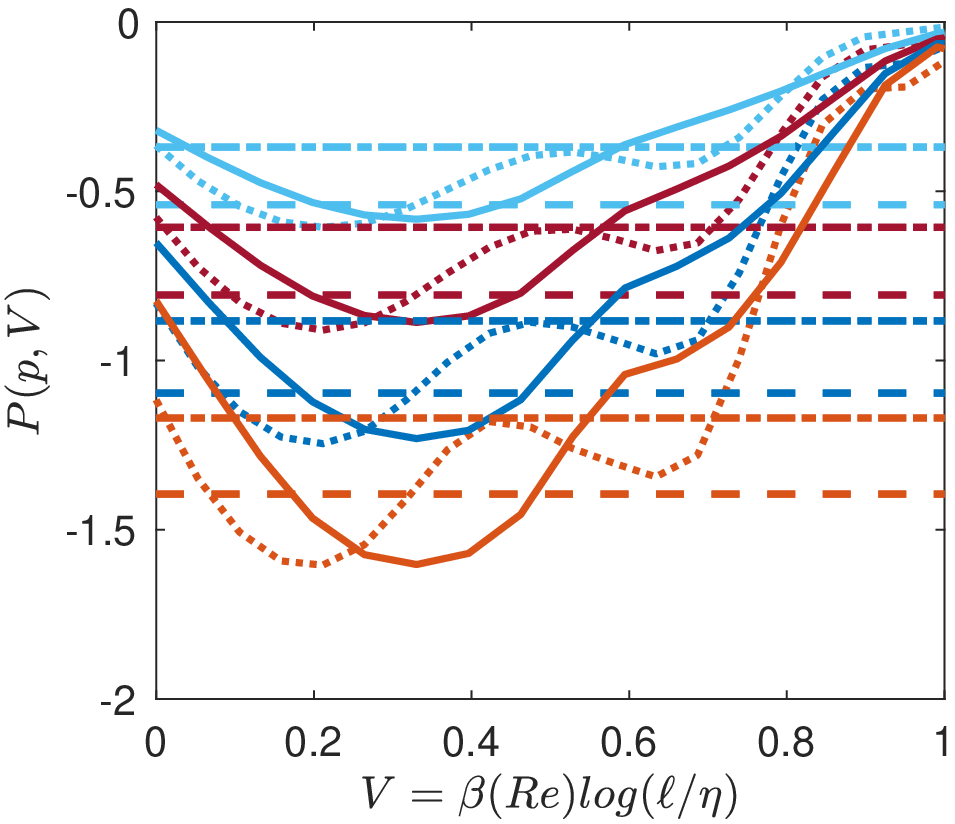}

	\caption{Same as figure \ref{fig:pressurelam} for $ \ R_{\lambda} = 90$ (line), $\ R_{\lambda} = 138$ (dotted line). Note the inflexion point appearing in the curves.}
\label{fig:pressureturb}
\end{figure}

Given our free energy, $F=\log(\tilde S_{3p})$, we can also compute the quantity conjugate to the volume, i.e. the multifractal pressure as: $P=\partial F/\partial V$. In the inertial range, 
where $\tilde S_p\sim \ell^{\tau(p)}$, we thus get $P=\tau(p)$, which only depends on the temperature. Outside the inertial range, $P$ has the meaning of a local scaling exponents also depends upon the scale, i.e., on the volume $V$ and on $N$ (Reynolds number). Using our universal functions derived in figure \ref{Fig:checkGUniv}, we can then compute empirically the multifractal pressure $P$ and see how it varies as a function of $T$, $V$ and $N$. It is provided in figure \ref{fig:pressurelam} for $ R_{\lambda} = 25$ and  $R_{\lambda} = 56$, and in figure  \ref{fig:pressureturb} for $ R_{\lambda} = 90$ and  $R_{\lambda} = 138$. We see that at low Reynolds number, the pressure decreases monotonically from the  dissipative range, reaches a lowest points  and then increases towards the largest scale. There is no clear flat plateau that would correspond to an "inertial" range. In contrast, at higher Reynolds number, a plateau   appears for $p=1$ to $p=4$ when going towards the largest scale, the value of the plateau corresponding to $\tau_{\text{DNS}}$. The plateau transforms into an inflection point for $p\ge 5$ making the derivative $\partial P/\partial V$ change sign. This is reminiscent of a phase transition occurring in the inertial range, with coexistence of two phases: one "laminar" and one "turbulent". We interpret such a phase transition as the result of the coexistence of region of flows with different H\"older exponents, with areas where the flow has been relaminarized due to the action of viscosity, because of the random character of the dissipative scale (see below). \

\section{Conclusion}
We have shown that a deep analogy exists between multifractal and classical thermodynamics. In this framework, one can derive from the usual velocity structure function an effective free energy that respects the classical extensivity properties, provided one uses a  number of degrees of freedom (given by $N=1/\beta(\text{Re}))$ that scales like $\log(R_\lambda$). This number is much smaller than the classical $N\sim \text{Re}^{9/4}$ that is associated with the number of nodes needed to discretize the Navier-Stokes equation down to the Kolmogorov scale. It would be interesting to see whether this number is also associated with the dimension of a suitable  "attractor of turbulence". 
Using the analogy, we also found the "multifractal" equation of state of turbulence, by computing the multifractal pressure $P=\partial F/\partial V$. We found that for large enough $R_\lambda$ and $p$ (the temperature), the system obeys a phase transition, with coexistence of phase like in the vapor-liquid transition. We interpret this phase transition as the result of the coexistence of region of flows with different H\"older exponents, with areas where the flow has been relaminarized due to the action of viscosity, because of the random character of the dissipative scale. We note that this kind of phenomenon has already been observed in the context of Lagrangian velocity increments, using the local scaling exponent $\zeta(p,\tau)=\mathrm{d}(\log(S_p(\tau)))/\mathrm{d}(\log(\tau))$ \cite{ArneodoLag08}. 
The phase transition is then associated with the existence of a fluctuating dissipative time scale. It has further been shown that  in a multifractal without fluctuating dissipative time scale, the local exponent decreases monotically from dissipative scale to large scale, implying a disappearance of the phase transition \cite{Biferale04}.\

\bibliography{thermoturb}

\end{document}